\documentclass[sigconf]{acmart}

\DeclareUnicodeCharacter{FF0C}{,}

\AtBeginDocument{%
  \providecommand\BibTeX{{%
    \normalfont B\kern-0.5em{\scshape i\kern-0.25em b}\kern-0.8em\TeX}}}

\usepackage{booktabs}
\usepackage{multirow}
\usepackage{threeparttable}
\usepackage{comment}
\usepackage{subcaption}

\usepackage{xspace}
\usepackage{amsmath}
\usepackage{amsfonts}
\usepackage{caption}

\usepackage{algorithm}
\usepackage{algorithmic}

\begin{document}

\title{Heteroscedastic Neural Surrogate Modeling for Robust and Rapid Bayesian Inference in Fusion Plasma Diagnostics}



\author{Liyun Zhang}
\affiliation{%
  \institution{GSFS, The University of Tokyo}
  \country{Japan}
}

\author{Naoya Mamada}
\affiliation{%
  \institution{GSFS, The University of Tokyo}
  \country{Japan}
}

\author{Kentaro Sakai}
\affiliation{%
  \institution{National Institute for Fusion Science}
  \country{Japan}
}

\author{Takeo Hoshi}
\affiliation{%
  \institution{National Institute for Fusion Science}
  \country{Japan}
}

\author{Toru Aonishi}
\affiliation{%
  \institution{GSFS, The University of Tokyo}
  \country{Japan}
}


\begin{abstract}
Bayesian inference via Markov Chain Monte Carlo (MCMC) provides effective parameter estimation, but its real-time application in complex physical systems is hindered by heavy computational bottlenecks and extreme sensitivity to statistical noise. We address this by proposing a neural-network-based probabilistic surrogate framework for rapid and robust MCMC inference. Using fusion plasma Thomson scattering diagnostics as a challenging, noise-dominated testbed, our approach employs a dual-head architecture to simultaneously estimate the expected physical emission spectrum and the channel-wise intrinsic measurement noise variance. By optimizing a Gaussian Negative Log-Likelihood (GNLL) objective, the learned aleatoric uncertainty dynamically buffers the sampler against pathological shot noise. Evaluations demonstrate that this surrogate framework achieves $>1500\times$ acceleration over exact physical forward models, while simultaneously reducing inference error (RMSE) by $>20\%$ compared to standard homoscedastic neural baselines, offering a highly promising paradigm for real-time physical analysis.
\end{abstract}












\maketitle

\section{Introduction}

\begin{figure}[t]
    \centering
    \includegraphics[width=\linewidth]{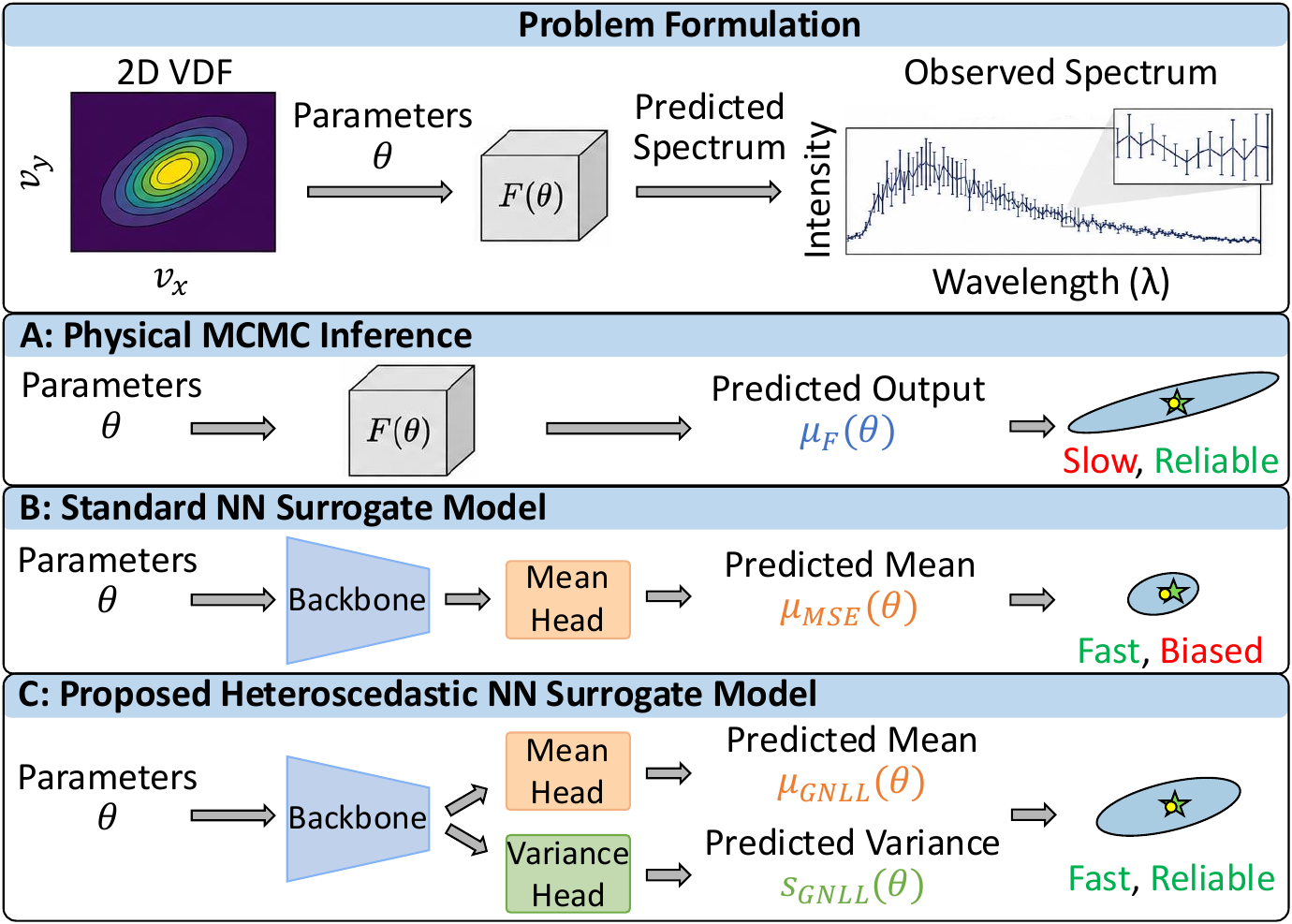}
    \caption{Comparative overview of MCMC inference in extreme-noise plasma diagnostics. Top: The forward model $F(\theta)$ maps latent VDFs to noisy spectra. Markers: green star (ground truth), yellow circle (prediction), and blue ellipse (posterior uncertainty). A: Physical MCMC yields theoretically exact posteriors but is computationally prohibitive and highly noise-sensitive (slender ellipse). B: Standard NN Surrogate (homoscedastic MSE loss) is fast but yields structurally biased and overconfident posteriors (circle far from star). C: Proposed Heteroscedastic Surrogate (GNLL loss) predicts data-dependent log-variance. This learned uncertainty acts as a topological smoother, penalizing overconfident proposals to enable fast and stable MCMC inference (ellipse's appropriate shape, nearly aligned circle and star).}
    \label{fig:comparison}
\end{figure}

Understanding the phase-space dynamics of particles is a cornerstone of modern physical sciences, particularly in controlled nuclear fusion. In magnetically confined plasmas, observing exact electron or ion velocity distribution functions (VDFs) is essential for diagnosing instabilities, heating efficiency, and non-Maxwellian anomalies \cite{sakai2026conceptual, kobayashi2026bayesian}. Advanced diagnostic systems, such as Thomson scattering, provide indirect spectral measurements of these VDFs \cite{kobayashi2026bayesian, rud2026bayesian}.

However, extracting the latent, high-dimensional VDFs from highly noisy, low-dimensional spectra constitutes a severely ill-posed inverse problem. Bayesian inference, particularly Markov Chain Monte Carlo (MCMC) sampling, is the gold standard for solving such problems, providing rigorous uncertainty quantification alongside accurate parameter estimation \cite{kobayashi2026bayesian, hammel2024machine} (Figure \ref{fig:comparison} A: the slender blue posterior ellipse captures the true physical degeneracy; the yellow predicted circle aligns with the green ground-truth star).

Despite its theoretical elegance, physical MCMC requires millions of sequential evaluations of the forward model to adequately explore the posterior distribution. In complex diagnostic environments, executing these first-principles simulations per step---which often involve intricate line integrations, atomic emission cross-sections, and instrumental broadenings---is computationally prohibitive, precluding rapid or near-real-time analysis. Replacing the expensive forward model with a Deep Neural Network (DNN) surrogate is a rapidly growing strategy to accelerate physical inference \cite{hammel2024machine, raissi2019physics}. However, standard surrogate models inherently assume homoscedastic (constant) observation noise by minimizing the Mean Squared Error (MSE) during training. 

This homoscedastic assumption critically breaks down in extreme scientific environments governed by photon statistics. For instance, in high-temperature and low-density plasmas, weak scattered photon signals cause observation noise to depend heavily on the input state and specific wavelength channels (i.e., heteroscedastic aleatoric uncertainty) \cite{sakai2026conceptual}. When embedded in an MCMC framework, a standard homoscedastic surrogate becomes structurally ``overconfident'' in these noisy regions. Lacking a mechanism to quantify varying noise, the surrogate incorrectly evaluates the Metropolis-Hastings likelihood, treating low-signal, high-noise channels with the same confidence as high-signal regions. 

Consequently, the sampler accepts physically invalid proposals, yielding heavily biased, uncalibrated posterior distributions (Figure \ref{fig:comparison} B: the collapsed blue ellipse tightly bounds a biased yellow predicted circle, significantly deviating from the green ground-truth star) that lose their physical validity.

To bridge the gap between high-speed surrogate modeling and reliable Bayesian inference in extreme noise environments, we propose a novel Heteroscedastic Neural Surrogate Model. Building upon aleatoric uncertainty quantification \cite{kendall2017what}, our dual-head architecture simultaneously outputs the predicted mean $\mu_{\text{GNLL}}(\theta)$ and a data-dependent log-variance $s_{\text{GNLL}}(\theta)$. By optimizing a Gaussian Negative Log-Likelihood (GNLL) objective, the network dynamically attenuates loss in high aleatoric uncertainty regions without explicit noise annotations. 

Crucially, when integrated into the MCMC likelihood, this formulation strictly penalizes overconfident proposals in noisy phase-space regions. This mechanism forces the sampler to properly weigh the reliability of different spectral channels, thereby restoring the physical validity and calibration of the inference while maintaining surrogate-level speed (Figure \ref{fig:comparison} C: the blue ellipse has the appropriate shape; the yellow predicted circle is nearly aligned with the green ground-truth star).

In this paper, we focus on the inverse problem of inferring velocity distribution functions from plasma diagnostic spectra to verify our approach. 
Our contributions are as follows:
\begin{itemize}
    \item \textbf{A novel heteroscedastic surrogate architecture for extreme-noise environments:} We propose a dual-head neural network that explicitly quantifies input-dependent aleatoric uncertainty. This establishes a rigorous probabilistic foundation for accelerating computationally expensive physical forward models without requiring explicit noise annotations.
    \item \textbf{A dynamic likelihood regularization mechanism for robust MCMC integration:} To overcome the extreme sensitivity of standard samplers to statistical noise, we directly embed the learned channel-wise log-variance into the Metropolis-Hastings evaluation. This mechanism acts as a topological smoother, successfully preventing sampler collapse and bypassing noise-induced local minima.
    \item \textbf{A highly efficient and accurate paradigm for physical diagnostics:} We empirically demonstrate that our framework achieves $>1500\times$ acceleration over exact physical MCMC. Supported by decoupled ablation studies, our method reduces inference RMSE by $>20\%$ compared to standard homoscedastic baselines, achieving physical robustness and well-calibrated uncertainty quantification.
\end{itemize}

\begin{figure*}[t]
    \centering
    \includegraphics[width=\linewidth]{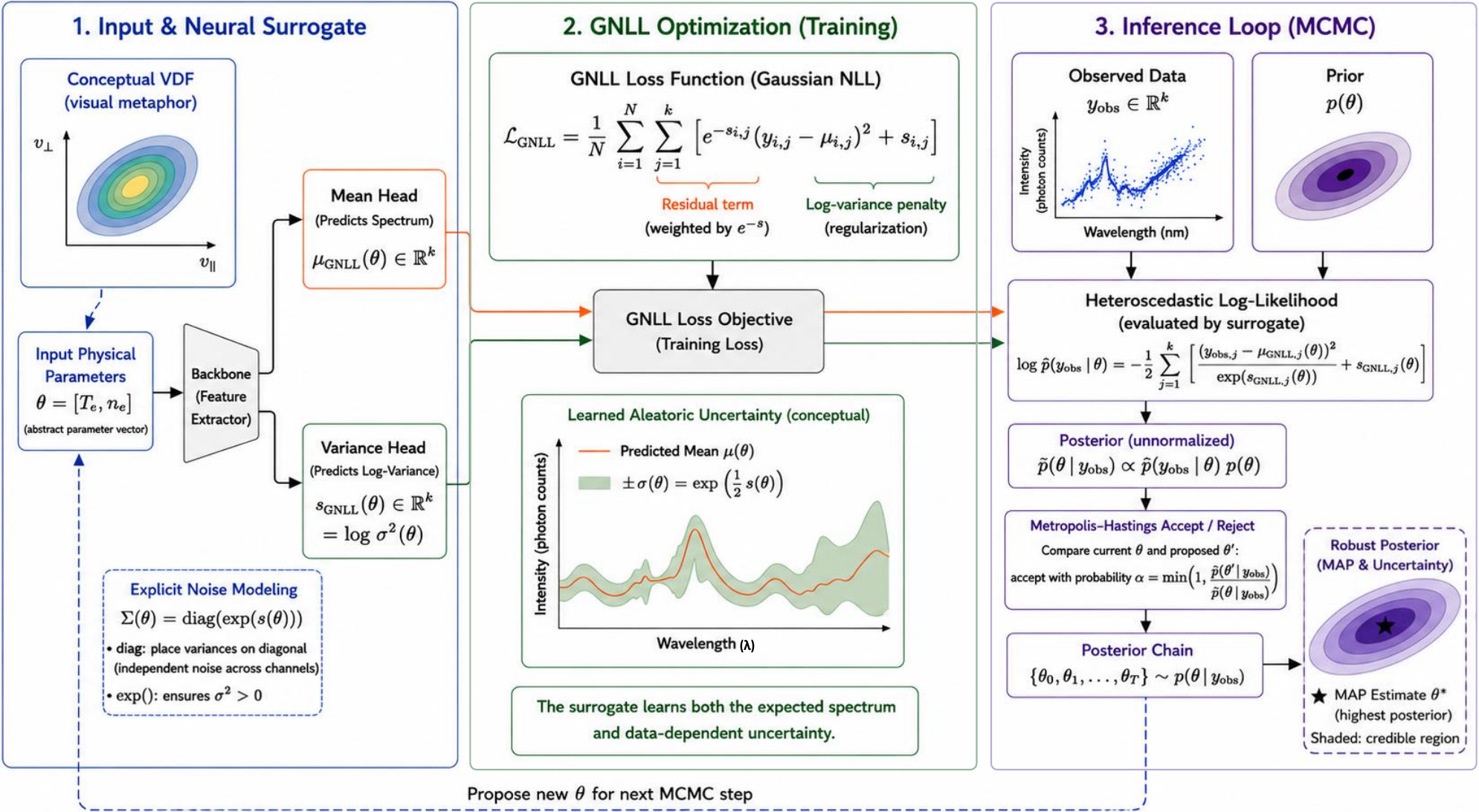}
    \caption{Workflow of the proposed heteroscedastic neural surrogate and its MCMC integration. (1) Neural Surrogate: A backbone processes abstract physical parameters $\theta$ (latent VDF) via a dual-head network, predicting expected spectrum $\mu_{\mathrm{GNLL}}(\theta)$ and channel-wise log-variance $s_{\mathrm{GNLL}}(\theta)$. (2) GNLL Optimization: The network is trained via Gaussian Negative Log-Likelihood (GNLL) loss, where learned aleatoric uncertainty attenuates residual penalties in high-noise regions. (3) MCMC Inference: The frozen surrogate evaluates the heteroscedastic log-likelihood. The learned variance regularizes the Metropolis-Hastings acceptance ratio, yielding a robust, physically consistent posterior.}
    \label{fig:method}
\end{figure*}

\section{Related Work}
\subsection{Bayesian Inference in Physical Diagnostics}
Bayesian inference provides a mathematically rigorous framework for physical inverse problems, which infer latent state variables from indirect and noisy observations \cite{stuart2010inverse}, enabling comprehensive uncertainty quantification (UQ) alongside point estimation. In plasma physics and fusion diagnostics, Bayesian methods have been widely adopted to infer velocity distribution functions (VDFs), temperature, and density profiles from integrated spectral data \cite{chilenski2015improved, kwak2024bayesian, fischer2003bayesian}. Markov Chain Monte Carlo (MCMC) algorithms (e.g., Metropolis-Hastings and Hamiltonian Monte Carlo) are standard workhorses for sampling these complex posteriors \cite{gelman1995bayesian, neal2011mcmc}. However, repetitive likelihood evaluations using computationally expensive physical forward models render traditional MCMC intractable for rapid analysis, especially in high-dimensional, extreme-noise environments \cite{cranmer2020frontiers, marzouk2007dimensionality}.

\subsection{Neural Surrogates for Forward Modeling}
To alleviate this computational bottleneck, deep neural networks (DNNs) are increasingly utilized as surrogate models to approximate complex physical simulations \cite{kasim2020building, karniadakis2021physics}. By training a neural network on a pre-computed dataset of physical parameters and their corresponding simulated observations, the expensive physical forward model can be replaced by a single, highly optimized forward pass of the DNN \cite{conrad2015accelerating, zhu2018bayesian}. This paradigm has achieved remarkable success across various physical domains \cite{raissi2019physics}. Despite these advancements, standard neural surrogates are predominantly trained using the Mean Squared Error (MSE) loss, which is statistically equivalent to assuming homoscedastic (constant) Gaussian noise across all data points \cite{goodfellow2016deep}. While effective for uniform noise, this assumption structurally fails in photon-starved diagnostic environments, leading to biased predictions and overconfident posterior evaluations when integrated into MCMC frameworks \cite{yang2021bpinns, wang2021understanding}.

\subsection{Uncertainty Quantification in Deep Learning}
Addressing the limitations of homoscedastic surrogates requires explicitly modeling data-dependent uncertainty \cite{abdar2021review}. In deep learning, uncertainty is categorized into epistemic uncertainty (model ignorance) and aleatoric uncertainty (inherent data noise) \cite{kendall2017what, hullermeier2021aleatoric}. While Bayesian Neural Networks and Monte Carlo Dropout are popular for epistemic uncertainty \cite{gal2016dropout, lakshminarayanan2017simple}, they do not capture heteroscedastic noise variations. To model heteroscedastic aleatoric uncertainty, Nix and Weigend \cite{nix1994estimating} pioneered the dual-head architecture, which Kendall and Gal \cite{kendall2017what} further formalized using a Gaussian Negative Log-Likelihood (GNLL) objective to attenuate loss dynamically \cite{seitzer2022pitfalls}. Our work bridges this gap for physical inverse problems by embedding a heteroscedastic surrogate within an MCMC loop, leveraging the predicted log-variance to restore physical calibration in extreme, data-dependent noise environments.

\section{Methodology}
\label{sec:method}

We formulate the plasma diagnostic inverse problem and demonstrate why homoscedastic surrogates collapse under extreme, data-dependent noise. We then detail our proposed heteroscedastic neural surrogate architecture and its robust integration into the MCMC inference framework.

\subsection{Problem Formulation and MCMC Inference}
Let $\theta \in \mathbb{R}^d$ denote the latent physical parameters (e.g., parameterizations of the highly non-linear velocity distribution function in a magnetically confined plasma). The observation relies on a computationally demanding physical forward model $F(\cdot)$, involving intricate line-of-sight integrations. The measured data is corrupted by stochastic noise:
\begin{equation}
    y_{\mathrm{obs}} = F(\theta) + \epsilon, \quad \epsilon \sim \mathcal{N}(0, \Sigma(\theta))
\end{equation}
where $y_{\mathrm{obs}} \in \mathbb{R}^k$ represents the observed diagnostic spectrum (e.g., discrete photon counts across specific wavelength channels). In photon-starved extreme environments, the noise $\epsilon$ is dominated by Poisson-like shot noise. Thus, the covariance matrix $\Sigma(\theta)$ is strictly heteroscedastic—strongly dependent on the input state and highly non-uniform across different spectral channels \cite{tarantola2005inverse}.

The objective of Bayesian inference is to estimate the posterior distribution $p(\theta | y_{\mathrm{obs}}) \propto p(y_{\mathrm{obs}} | \theta) p(\theta)$, where $p(\theta)$ embodies our prior physical knowledge and $p(y_{\mathrm{obs}} | \theta)$ is the likelihood function. Due to the high dimensionality and non-linearity of $F(\theta)$, the posterior is analytically intractable. Thus, the Metropolis-Hastings (MH) algorithm \cite{hastings1970monte} is employed to sample the posterior space. At each time step $t$, a candidate state $\theta'$ is drawn from a proposal distribution $q(\theta' | \theta_t)$ and accepted with probability:
\begin{equation}
    \alpha = \min \left( 1, \frac{p(y_{\mathrm{obs}} | \theta') p(\theta') q(\theta_t | \theta')}{p(y_{\mathrm{obs}} | \theta_t) p(\theta_t) q(\theta' | \theta_t)} \right)
\end{equation}
Executing this MH step requires evaluating the expensive physical simulator $F(\theta)$ millions of times, rendering rapid physical analysis strictly intractable. To resolve this, our fully integrated surrogate-accelerated MH sampling framework will be systematically summarized in Algorithm \ref{alg:mcmc} in the \textit{Robust MCMC Integration} section.

\subsection{Failure Mode of Homoscedastic Surrogates}
To bypass the computational bottleneck, standard approaches replace $F(\theta)$ with a fast neural surrogate $\mu_{\mathrm{MSE}}(\theta)$, typically trained by minimizing the Mean Squared Error (MSE) over a pre-computed dataset $\mathcal{D} = \{ (\theta_i, y_i) \}_{i=1}^N$:
{\small
\begin{equation}
    \mathcal{L}_{\mathrm{MSE}} = \frac{1}{N} \sum_{i=1}^N \left\| y_i - \mu_{\mathrm{MSE}}(\theta_i) \right\|^2
\end{equation}}
Crucially, from a maximum likelihood perspective, minimizing the MSE is mathematically isomorphic to assuming that the observational noise follows an isotropic, homoscedastic Gaussian distribution, $\Sigma = \sigma^2 I$ \cite{goodfellow2016deep}. When this standard surrogate is embedded within the MCMC framework, the approximate log-likelihood is uniformly evaluated as:
\begin{equation}
    \log p(y_{\mathrm{obs}} | \theta) \approx -\frac{1}{2\sigma^2} \left\| y_{\mathrm{obs}} - \mu_{\mathrm{MSE}}(\theta) \right\|^2 + C
\end{equation}

This rigid homoscedastic assumption is fatal in diagnostic environments governed by photon statistics. In extremely low-signal spectral channels, massive shot noise causes the residual $(y_{\mathrm{obs}} - \mu_{\mathrm{MSE}})^2$ to fluctuate wildly. The uniform scalar variance $\sigma^2$ strictly penalizes these fluctuations, misinterpreting them as fundamental physical mismatches rather than inherent data uncertainties. Consequently, noise-induced gradients mislead the MCMC sampler, causing it to reject valid proposals and collapse into a pathologically biased, overconfident posterior.

\subsection{Proposed Heteroscedastic Surrogate Optimization}
To capture the data-dependent noise landscape without explicit annotations, we propose a novel dual-head heteroscedastic architecture. A Multi-Layer Perceptron (MLP) backbone (four 256-neuron hidden layers with ReLU activations) extracts physics-aware features from the input $\theta$. This lightweight design reduces the forward pass to microsecond-level matrix multiplications before bifurcating into two task-specific predictive heads:

\textbf{Mean Head:} Predicts the expected physical spectrum, denoted as $\mu_{\mathrm{GNLL}}(\theta) \in \mathbb{R}^k$.

\textbf{Variance Head:} Predicts a high-dimensional, \textit{channel-wise} data-dependent log-variance vector, $s_{\mathrm{GNLL}}(\theta) = \log \sigma^2_{\mathrm{GNLL}}(\theta) \in \mathbb{R}^k$. Unlike standard global scalar variance models, this fine-grained parameterization is structurally indispensable to capture the highly localized, non-uniform photon shot-noise spikes at spectral edges.

Predicting the log-variance ($s$) rather than absolute variance ($\sigma^2$) avoids constrained optimization and guarantees numerical stability (strictly positive variance) \cite{kendall2017what}. The network is optimized end-to-end via the Gaussian Negative Log-Likelihood (GNLL) objective:
{\small
\begin{equation}
    \mathcal{L}_{\mathrm{GNLL}} = \frac{1}{N} \sum_{i=1}^N \sum_{j=1}^k \frac{1}{2} \left[ \exp(-s_{i,j}) \left( y_{i,j} - \mu_{i,j} \right)^2 + s_{i,j} \right]
\end{equation}}
where $j$ indexes individual spectral channels. The GNLL objective acts as a powerful self-supervised attenuation mechanism. In phase-space regions characterized by extreme shot noise, the network is naturally penalized for making precise point predictions. To prevent the loss from exploding due to large $(y - \mu)^2$ residuals, the network automatically learns to predict a large $s_{i,j}$, which elegantly down-weights the noisy observation and regularizes the predictive uncertainty.

\begin{algorithm}[!t]
\caption{Heteroscedastic Surrogate-Accelerated MCMC}
\label{alg:mcmc}
\begin{algorithmic}[1]
\REQUIRE Observed data $y_{\mathrm{obs}}$, Pre-trained dual-head surrogate $(\mu_{\mathrm{GNLL}}, s_{\mathrm{GNLL}})$, Initial state $\theta_0$, Prior $p(\theta)$, Proposal distribution $q(\cdot|\cdot)$, Steps $T$
\FOR{$t = 0$ to $T-1$}
    \STATE Draw candidate $\theta' \sim q(\theta' | \theta_t)$
    \STATE Forward pass through the frozen neural surrogate: \\
    \hspace{1cm}$\mu', s' = \mathrm{SurrogateForward}(\theta')$
    \STATE Evaluate heteroscedastic log-likelihood: \\
    \hspace{1cm}$\log \hat{p}(y_{\mathrm{obs}} | \theta') = \sum_{j} \left( -\frac{(y_{\mathrm{obs},j} - \mu'_j)^2}{2\exp(s'_j)} - \frac{s'_j}{2} \right)$
    \STATE Compute Metropolis-Hastings ratio in log domain: \\
    \hspace{1cm}$\begin{aligned} 
        \log \alpha &= \log \hat{p}(y_{\mathrm{obs}} | \theta') + \log p(\theta') \\
        &\quad - \log \hat{p}(y_{\mathrm{obs}} | \theta_t) - \log p(\theta_t)
    \end{aligned}$
    \STATE Draw uniform random variable $u \sim \mathrm{Uniform}(0, 1)$
    \IF{$\log u < \log \alpha$}
        \STATE Accept proposal: \\
        \hspace{1cm}$\theta_{t+1} = \theta'$
    \ELSE
        \STATE Reject, stay at current state: \\
        \hspace{1cm}$\theta_{t+1} = \theta_t$
    \ENDIF
\ENDFOR
\item[\textbf{Ensure:}] Chain $\{\theta_0, \dots, \theta_T\}$ representing the robust posterior $p(\theta | y_{\mathrm{obs}})$
\end{algorithmic}
\end{algorithm}

\subsection{Robust MCMC Integration}
During the inference phase, both the predicted mean and learned uncertainty actively dictate the Metropolis-Hastings acceptance probability. Instead of a uniform MSE distance, the heteroscedastic log-likelihood evaluated by our surrogate is dynamically adjusted per channel:
{\small
\begin{equation}
\begin{split}
    \log \hat{p}(y_{\mathrm{obs}} | \theta) = &-\frac{1}{2} \sum_{j=1}^k \bigg[ \frac{( y_{\mathrm{obs}, j} - \mu_{\mathrm{GNLL}, j}(\theta) )^2}{\exp(s_{\mathrm{GNLL}, j}(\theta))} \\
    & + s_{\mathrm{GNLL}, j}(\theta) \bigg]
\end{split}
\label{eq:hetero_likelihood}
\end{equation}}
This formulation explicitly penalizes the sampler for becoming overconfident in regions where the surrogate indicates high aleatoric uncertainty ($\exp(s_{\mathrm{GNLL}})$ is large). As detailed in Algorithm \ref{alg:mcmc}, this dynamically adjusted log-likelihood is continuously evaluated at each MCMC step. By doing so, the surrogate entirely replaces the computationally prohibitive simulator $F(\theta)$ while inherently guarding the Markov chain against noise-induced collapse.

\begin{figure*}[t]
    \centering
    \includegraphics[width=\textwidth]{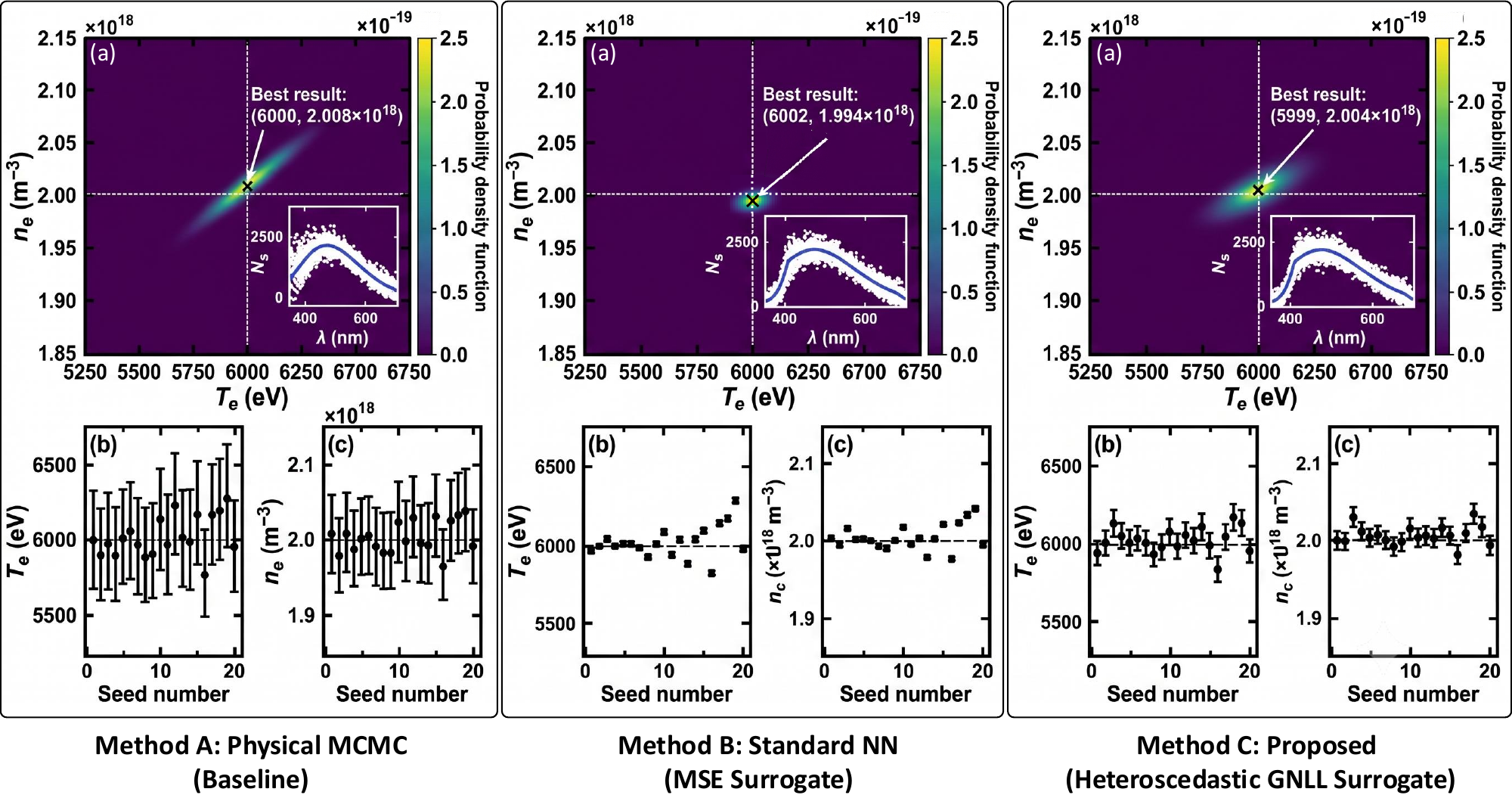}
    \caption{Qualitative comparison of 2D Posterior Probability Density Functions (a: best seeds with minimal absolute $T_e$ error, marked $\boldsymbol{\dagger}$ in Table \ref{tab:quantitative_results}) and parameter trace plots across 20 independent seeds (b, c). Method A: yields precise `best-case' estimates, though 95\% credible intervals remain widely dispersed due to extreme noise sensitivity. Method B: collapses into a biased and overconfident geometry (almost vanishing error bars) by overfitting localized shot noise. Method C: Our proposed GNLL Surrogate acts as a topological smoother, bypassing noise-induced local minima. Its 95\% credible intervals consistently encompass ground-truth values, ensuring the estimated uncertainty aligns with actual estimation errors under extreme noise.}
    \label{fig:qualitative_posterior}
\end{figure*}

\subsection{Theoretical Advantages and Computational Cost}
Integrating the heteroscedastic surrogate (Operational loop detailed in Algorithm \ref{alg:mcmc}) yields two profound advantages.  

First, in computational complexity, standard MCMC evaluates a numerical solver with time complexity $\mathcal{O}(T \cdot \mathcal{C}_{F})$, where $\mathcal{C}_{F}$ is the massive cost of the physical simulator. Replacing $F(\theta)$ with a neural network reduces this to pure GPU matrix multiplications, $\mathcal{O}(T \cdot \mathcal{C}_{\mathrm{NN}})$ (where $\mathcal{C}_{\mathrm{NN}} \ll \mathcal{C}_{F}$), facilitating $>1500\times$ acceleration in end-to-end inference time as empirically validated in our subsequent experiments. 

Second, in statistical calibration, including the data-dependent variance $\exp(s')$ in the likelihood denominator actively guides the sampler's geometric random walk. This prevents the chain from trapping in localized, noise-induced likelihood peaks, ensuring the posterior remains unbiased, highly robust, and physically meaningful even in extreme-noise diagnostic scenarios.

\begin{table*}[t]
\centering
\small
\resizebox{\textwidth}{!}{%
\begin{tabular}{l|cc|cc|cc}
\toprule
\multirow{2}{*}{Test ID} & \multicolumn{2}{c|}{Method A (Physical MCMC)} & \multicolumn{2}{c|}{Method B (Standard MSE)} & \multicolumn{2}{c}{Method C (Proposed GNLL)} \\ 
 & $T_e$ (eV) & $n_e (\times 10^{18} \text{ m}^{-3})$ & $T_e$ (eV) & $n_e (\times 10^{18} \text{ m}^{-3})$ & $T_e$ (eV) & $n_e (\times 10^{18} \text{ m}^{-3})$ \\ \midrule
Seed 01 & $6000.18^{\boldsymbol{\dagger}}$ & $2.008^{\boldsymbol{\dagger}}$ & $5975.22$ & $2.004$ & $5948.87$ & $1.998$ \\
Seed 02 & $5895.83$ & $1.979$ & $6002.32^{\boldsymbol{\dagger}}$ & $1.994^{\boldsymbol{\dagger}}$ & $6017.00$ & $1.997$ \\
Seed 03 & $5984.37$ & $2.007$ & $6051.74$ & $2.019$ & $6139.09$ & $2.032$ \\
Seed 04 & $5894.42$ & $1.987$ & $6002.89$ & $2.001$ & $6060.64$ & $2.013$ \\
Seed 05 & $6015.62$ & $2.003$ & $6022.20$ & $2.002$ & $6020.44$ & $2.002$ \\
Seed 06 & $6062.57$ & $2.006$ & $6022.95$ & $2.000$ & $6043.94$ & $2.011$ \\
Seed 07 & $5968.76$ & $1.992$ & $5991.02$ & $1.990$ & $6015.72$ & $1.998$ \\
Seed 08 & $5879.24$ & $1.984$ & $5931.52$ & $1.986$ & $5942.97$ & $1.985$ \\
Seed 09 & $5896.20$ & $1.983$ & $6019.44$ & $2.000$ & $5992.48$ & $1.997$ \\
Seed 10 & $6146.50$ & $2.027$ & $6093.20$ & $2.020$ & $6085.99$ & $2.012$ \\
Seed 11 & $5967.27$ & $1.997$ & $5949.18$ & $1.996$ & $5993.97$ & $2.001$ \\
Seed 12 & $6233.58$ & $2.031$ & $6047.51$ & $2.003$ & $6067.07$ & $2.002$ \\
Seed 13 & $6014.16$ & $1.996$ & $5889.76$ & $1.977$ & $6033.08$ & $2.000$ \\
Seed 14 & $5987.02$ & $1.994$ & $6049.30$ & $2.001$ & $6118.79$ & $2.014$ \\
Seed 15 & $6162.24$ & $2.033$ & $6103.89$ & $2.022$ & $5999.16^{\boldsymbol{\dagger}}$ & $2.002^{\boldsymbol{\dagger}}$ \\
Seed 16 & $5765.62$ & $1.964$ & $5832.79$ & $1.975$ & $5843.39$ & $1.976$ \\
Seed 17 & $6159.86$ & $2.027$ & $6154.28$ & $2.024$ & $6057.17$ & $2.010$ \\
Seed 18 & $6189.49$ & $2.036$ & $6181.48$ & $2.034$ & $6178.45$ & $2.034$ \\
Seed 19 & $6292.66$ & $2.039$ & $6296.60$ & $2.043$ & $6142.83$ & $2.016$ \\
Seed 20 & $5952.12$ & $1.993$ & $5983.41$ & $1.992$ & $5963.85$ & $1.987$ \\ 
\midrule
Ensemble Std & $135.71$ & $0.0214$ & $103.43$ & $0.0179$ & $\mathbf{78.88}$ & $\mathbf{0.0141}$ \\
RMSE to Truth & $134.32$ & $0.0213$ & $105.19$ & $0.0179$ & $\mathbf{83.76}$ & $\mathbf{0.0144}$ \\
\midrule
Inference Time & \multicolumn{2}{c|}{$\sim 11.1$ hours / PDF} & \multicolumn{2}{c|}{$\mathbf{20.3 \text{ sec}}$ / PDF} & \multicolumn{2}{c}{$26.1 \text{ sec}$ / PDF} \\ \bottomrule
\end{tabular}%
}
\caption{Comprehensive MCMC inference results across 20 independent extreme-noise blind tests (Target Truth: $T_e = 6000 \text{ eV}, n_e = 2.000 \times 10^{18} \text{ m}^{-3}$). The summary metrics---Ensemble Std, RMSE to Truth, and Inference Time per posterior PDF---quantitatively demonstrate that the proposed GNLL surrogate (Method C) achieves $\sim 1500\times$ acceleration over the physical MCMC (Method A) while surpassing the standard MSE (Method B) in both accuracy (RMSE) and stability (Std). Values marked with a dagger ($\boldsymbol{\dagger}$) denote seeds visualized in Figure \ref{fig:qualitative_posterior} (a), chosen for minimum absolute $T_e$ error.}
\label{tab:quantitative_results}
\end{table*}

\section{Experiments}
\label{sec:experiments}
We evaluate our method against both the physical MCMC baseline \cite{sakai2026conceptual} and standard neural surrogates \cite{hammel2024machine} across accuracy, robustness, and computational cost.

\textbf{Implementation Details.}
Models were implemented in PyTorch and trained on an NVIDIA A100 GPU using the Adam optimizer (learning rate: $1 \times 10^{-3}$, batch size: $2048$, epochs: $100$). For inference, Bayesian sampling utilized the affine-invariant \texttt{emcee} sampler \cite{foreman2012emcee} with $32$ walkers for $2000$ steps (the initial $500$ discarded as burn-in).

\textbf{Dataset and Preprocessing.}
We utilized $1,340,000$ Thomson scattering spectra generated via the official physical forward model, spanning an electron temperature ($T_e$) range of $100.0 \text{ eV}$ to $6780.0 \text{ eV}$; the setup follows the cited configuration with $k=2560$ channels \cite{sakai2026conceptual}. To replicate extreme photon-starved conditions, simulated counts underwent an instrumental efficiency inverse-correction. Because detection efficiency drops drastically at spectral edges ($<400 \text{ nm}$ and $>600 \text{ nm}$), this induces severe, non-uniform heteroscedastic noise across 2560 channels.

\textbf{Baselines.}
We benchmark our approach (Method C: Proposed GNLL Surrogate) against:
(1) Method A (Physical MCMC): The domain-standard approach utilizing Population Annealing Monte Carlo (PAMC) \cite{sakai2026conceptual}, requiring 200,000 expensive physical likelihood evaluations per posterior.
(2) Method B (Standard MSE Surrogate): Uses an identical MLP backbone trained via MSE loss, implicitly assuming uniform homoscedastic noise ($\Sigma = \sigma^2 I$) and serves as the conventional neural network baseline.

\textbf{Metrics.}
Performance is evaluated across an ensemble of $N=20$ independent MCMC blind tests (different noise seeds) using three metrics:
(1) Statistical Robustness (Ensemble Std): Unbiased sample standard deviation, $\text{Std}(x) = \sqrt{\frac{1}{N-1}\sum_{i=1}^{N}(x_i - \bar{x})^2}$, strictly measuring inference stability against noise-induced fluctuations.
(2) Overall Accuracy (RMSE): Root Mean Square Error, $\text{RMSE}(x) = \sqrt{\frac{1}{N}\sum_{i=1}^{N}(x_i - x^*)^2}$, simultaneously penalizing absolute bias from the truth and wild variance.
(3) Computational Efficiency: End-to-end MCMC execution time required per 2D posterior probability density function (PDF).

\begin{table*}[t]
\centering
\small
\begin{tabular}{llccccc}
\toprule
Variant & Variance Modeling & MCMC Likelihood & Std ($T_e$) & Std ($n_e$) & RMSE ($T_e$) & RMSE ($n_e$) \\ 
\midrule
Standard MSE & None & Uniform & $103.43$ & $0.0179$ & $105.19$ & $0.0179$ \\
Global GNLL & Global Scalar & Uniform & $90.51$ & $0.0145$ & $88.29$ & $0.0146$ \\
Decoupled GNLL & Channel Vector & Uniform & $87.48$ & $0.0143$ & $85.84$ & $\mathbf{0.0144}$ \\
Proposed GNLL & \textbf{Channel Vector} & \textbf{Data-Dependent} & $\mathbf{78.88}$ & $\mathbf{0.0141}$ & $\mathbf{83.76}$ & $\mathbf{0.0144}$ \\ 
\bottomrule
\end{tabular}
\caption{Ablation study decoupling the architectural and inferential components of our framework. We evaluate the necessity of channel-wise variance modeling and its explicit integration into the MCMC likelihood evaluation. Metrics isolate the inference precision (unbiased Std and RMSE), with units being eV for $T_e$ and $\times 10^{18} \text{ m}^{-3}$ for $n_e$.}
\label{tab:ablation_loss}
\end{table*}

\subsection{Qualitative Analysis}
We map the 2D posterior PDFs of the best-performing test seed (minimum absolute $T_e$ error, marked $\boldsymbol{\dagger}$ in Table \ref{tab:quantitative_results}) in Figure \ref{fig:qualitative_posterior} (a), alongside MCMC parameter trace plots across the 20-seed ensemble in Figure \ref{fig:qualitative_posterior} (b, c).

\textbf{Method A (Physical MCMC)} illustrates the exact physical posterior topology via exact physical simulator \cite{sakai2026conceptual}. The diagonally elongated credible region 2D contour accurately reflects the inherent physical degeneracy between temperature and density. While Method A recovers a near-perfect estimate under this optimal seed, trace plots across all 20 independent seeds show wide, highly fluctuant exploration variance. These wide error bars (95\% credible intervals) correctly capture the severe true physical uncertainty mandated by the extreme-noise environment.

\textbf{Method B (Standard MSE Surrogate)} exposes the critical flaw of the Standard NN. Because the MSE loss forces a rigid, uniform variance assumption across all 2560 channels, the surrogate is coerced into overfitting the massive, randomly distributed shot-noise spikes at the low-efficiency spectral edges. These artificial gradient cliffs deceive the Metropolis-Hastings criterion, severely trapping the sampler in noise-induced local minima. This yields a pathologically \textit{overconfident} posterior—visually evident as an unnaturally concentrated tiny dot in the 2D PDF and vanishingly small error bars in the trace plots—completely failing to capture the physical credible region.

\textbf{Method C (Heteroscedastic GNLL Surrogate)} validates our proposed framework. Guided by the GNLL objective, the Variance Head autonomously identifies regions of severe aleatoric noise. During MCMC, this learned variance acts as a dynamic likelihood regularizer, safely attenuating pathological noise penalties. Trace plots across the 20 seeds demonstrate healthy exploration variance, with 95\% credible intervals consistently encompassing ground-truth values, indicating estimated uncertainty is compatible with the observed estimation error under extremely noisy conditions. The framework recovers highly stable point estimates and well-calibrated error bars, suggesting that it provides robust uncertainty quantification without a noticeable loss of computational efficiency relative to standard surrogates.

\subsection{Quantitative Analysis}
Table \ref{tab:quantitative_results} enumerates the inferred state vectors $(T_e, n_e)$ across 20 extreme-noise seeds, corroborating the trace dynamics observed in Figure \ref{fig:qualitative_posterior}. We benchmark three critical dimensions: computational efficiency, physical robustness, and uncertainty quantification.

\textbf{Efficiency \& Physical Robustness.}
Method A serves as the theoretical gold standard, but executing $200,000$ PAMC steps requires $\sim 11.1$ hours \cite{sakai2026conceptual}. Both neural surrogates accelerate inference to under $30$ seconds (a $>1500\times$ speedup). 

However, acceleration demands stability. Under extreme shot noise, Method B swings violently between $5832.79 \text{ eV}$ and $6296.60 \text{ eV}$. We rigorously quantify this using the unbiased \textit{Ensemble Std} and \textit{RMSE}. By leveraging the GNLL objective to dynamically down-weight noisy spectral edges, our proposed Method C compresses the $T_e$ standard deviation to $78.88 \text{ eV}$ (vs. Method B's $103.43 \text{ eV}$) and reduces the RMSE to $83.76 \text{ eV}$ (a $>20\%$ improvement over Method B).  

Notably, all methods estimate $n_e$ with relative stability. Physically, $n_e$ is proportional to the total spectral integral, making it inherently resilient to localized noise. Conversely, $T_e$ strictly depends on noise-sensitive spectral broadening \cite{sakai2026conceptual}, making Method C's dynamic variance attenuation essential to prevent catastrophic overfitting of tail artifacts.

\textbf{Discussion: Surrogate Regularization vs. Physical Overfitting.}
An intriguing observation from Table \ref{tab:quantitative_results} is that the proposed neural surrogate (Method C) achieves a lower aggregate RMSE and tighter Ensemble Std than the exact physical forward model (Method A), which possesses the highest theoretical accuracy bounds and can yield precise estimates under favorable noise seeds (see Figure \ref{fig:qualitative_posterior} (a)).

However, in photon-starved, extreme-noise regimes, the physical MCMC is overly sensitive to massive stochastic spikes, leading to significant trace fluctuations across the 20-seed ensemble. In contrast, having optimized its weights over 1.34 million diverse parameter-spectrum pairs, the neural surrogate implicitly encodes the smooth structural manifold of the physical data. By coupling this with the GNLL variance attenuation, Method C acts as a topological smoother that intelligently ignores localized extreme noise rather than overfitting it. This data-driven regularization dramatically improves the conditioning of the inverse problem, trading microscopic theoretical fidelity for massive statistical stability and robustness.

\subsection{Ablation Study}
To validate our framework's rationality, we isolate the neural architecture and probabilistic inference mechanisms to evaluate all variants across both $T_e$ and $n_e$ parameters, summarized in Table \ref{tab:ablation_loss}.

\textbf{The Granularity of Variance Modeling.} 
We first ablate heteroscedasticity. Stripping the architecture to a standard MSE Baseline yields an unacceptable $T_e$ RMSE of $105.19 \text{ eV}$. Introducing a Global GNLL variant (predicting a single log-variance scalar) improves $T_e$ RMSE to $88.29 \text{ eV}$ and stabilizes $n_e$, but fails to capture highly localized photon shot-noise spikes at the spectral edges, proving fine-grained, \textit{channel-wise} variance modeling is structurally indispensable.

\textbf{The Necessity of MCMC Integration.} 
We then isolate the learned uncertainty's role during sampling. In the Decoupled GNLL variant, the network is trained with the channel-wise GNLL objective, but the MCMC loop forces a standard uniform likelihood. Although GNLL training provides implicit regularization (reducing $T_e$ RMSE to $85.84 \text{ eV}$), the sampler lacks dynamic data-dependent guidance and still occasionally falls into noise-induced local minima. 

\textbf{Full Framework.} 
The full framework explicitly embeds the learned channel-wise variance into the MCMC likelihood denominator (Equation \ref{eq:hetero_likelihood}). Results confirm this integration is essential for optimal precision, validating that the heteroscedastic surrogate acts as a topological smoother to dynamically buffer the posterior landscape against extreme stochastic fluctuations.

\section{Conclusion}
\label{sec:conclusion}
In this work, we proposed a heteroscedastic surrogate-accelerated MCMC framework to resolve the computational bottlenecks and noise-induced instabilities in Bayesian plasma diagnostics. By embedding a dual-head GNLL architecture directly into the Metropolis-Hastings likelihood evaluation, our approach autonomously mitigates pathological shot noise. This framework accelerates exact posterior inference by $>1500\times$ while significantly surpassing standard baselines in physical robustness and accuracy. Future work will extend this paradigm to multi-diagnostic sensor fusion for real-time plasma control.

\bibliographystyle{ACM-Reference-Format}
\bibliography{sample-base}

@String{Computing = "Computing" }

@String{Computer = "{IEEE} Computer" }

@String{Springer = "Springer-Verlag" }

@article{kendall2017what,
  title={What uncertainties do we need in bayesian deep learning for computer vision?},
  author={Kendall, Alex and Gal, Yarin},
  journal={Advances in neural information processing systems},
  volume={30},
  year={2017}
}

@article{raissi2019physics,
  title={Physics-informed neural networks: A deep learning framework for solving forward and inverse problems involving nonlinear partial differential equations},
  author={Raissi, Maziar and Perdikaris, Paris and Karniadakis, George E},
  journal={Journal of Computational physics},
  volume={378},
  pages={686--707},
  year={2019},
  publisher={Elsevier}
}

@article{hammel2024machine,
  title={Machine learning assisted bayesian inference of mix and hot-spot conditions in NIF implosions},
  author={Hammel, BA and Hammel, BD and Scott, HA and Peterson, J Luc},
  journal={High Energy Density Physics},
  volume={50},
  pages={101077},
  year={2024},
  publisher={Elsevier}
}

@article{kobayashi2026bayesian,
  title={Bayesian modeling of charge exchange recombination spectroscopy in fusion plasmas},
  author={Kobayashi, T and Hoshi, T and Nakano, A and Yoshinuma, M and Ida, K},
  journal={Review of Scientific Instruments},
  volume={97},
  number={6},
  year={2026},
  publisher={AIP Publishing}
}

@article{rud2026bayesian,
  title={Bayesian velocity-space tomography with collision-and charge-exchange-physics prior from fast-ion D-alpha measurements at TCV with uncertainty quantification},
  author={Rud, M and Dong, Y and Everink, JM and J{\"a}rleblad, H and J{\o}rgensen, JS and Kernel, Q and Madsen, B and Podesta, M and Valentini, A and Salewski, M and others},
  journal={Plasma Physics and Controlled Fusion},
  volume={68},
  number={1},
  pages={015035},
  year={2026},
  publisher={IOP Publishing}
}

@article{sakai2026conceptual,
  title={Conceptual design of Thomson scattering system with high wavelength resolution in magnetically confined plasmas for electron phase-space measurements},
  author={Sakai, Kentaro and Tomita, Kentaro and Hoshi, Takeo and Nakano, Akito and Goto, Motoshi and Nagaoka, Kenichi and Yasuhara, Ryo},
  journal={Plasma Physics and Controlled Fusion},
  volume={68},
  number={1},
  pages={015034},
  year={2026},
  publisher={IOP Publishing}
}

@article{chilenski2015improved,
  title={Improved profile fitting and quantification of uncertainty in experimental measurements of impurity transport coefficients using Gaussian process regression},
  author={Chilenski, MA and Greenwald, M and Marzouk, Y and Howard, NT and White, AE and Rice, JE and Walk, JR},
  journal={Nuclear Fusion},
  volume={55},
  number={2},
  pages={023012},
  year={2015},
  publisher={IOP Publishing}
}

@article{kwak2024bayesian,
  title={Bayesian modelling of multiple plasma diagnostics at Wendelstein 7-X},
  author={Kwak, Sehyun and Hoefel, U and Krychowiak, M and Langenberg, A and Svensson, J and Trimino Mora, H and Ghim, Y-C and W7-X Team},
  journal={Nuclear Fusion},
  volume={64},
  number={10},
  pages={106022},
  year={2024},
  publisher={IOP Publishing}
}

@book{gelman1995bayesian,
  title={Bayesian data analysis},
  author={Gelman, Andrew and Carlin, John B and Stern, Hal S and Rubin, Donald B},
  year={1995},
  publisher={Chapman and Hall/CRC}
}

@article{cranmer2020frontiers,
  title={The frontier of simulation-based inference},
  author={Cranmer, Kyle and Brehmer, Johann and Louppe, Gilles},
  journal={Proceedings of the National Academy of Sciences},
  volume={117},
  number={48},
  pages={30055--30062},
  year={2020},
  publisher={National Academy of Sciences}
}

@article{kasim2020building,
  title={Building high accuracy emulators for scientific simulations with deep neural architecture search},
  author={Kasim, Muhammad Firmansyah and Watson-Parris, Duncan and Deaconu, Lucia and Oliver, Sophy and Hatfield, P and Froula, Dustin H and Gregori, Gianluca and Jarvis, Matt and Khatiwala, Samar and Korenaga, Jun and others},
  journal={Machine Learning: Science and Technology},
  volume={3},
  number={1},
  pages={015013},
  year={2022},
  publisher={IOP Publishing}
}

@article{conrad2015accelerating,
  title={Accelerating asymptotically exact MCMC for computationally intensive models via local approximations},
  author={Conrad, Patrick R and Marzouk, Youssef M and Pillai, Natesh S and Smith, Aaron},
  journal={Journal of the American Statistical Association},
  volume={111},
  number={516},
  pages={1591--1607},
  year={2016},
  publisher={Taylor \& Francis}
}

@article{yang2021bpinns,
  title={B-PINNs: Bayesian physics-informed neural networks for forward and inverse PDE problems with noisy data},
  author={Yang, Liu and Meng, Xuhui and Karniadakis, George Em},
  journal={Journal of Computational Physics},
  volume={425},
  pages={109913},
  year={2021},
  publisher={Elsevier}
}

@inproceedings{nix1994estimating,
  title={Estimating the mean and variance of the target probability distribution},
  author={Nix, David A and Weigend, Andreas S},
  booktitle={Proceedings of 1994 ieee international conference on neural networks (ICNN'94)},
  volume={1},
  pages={55--60},
  year={1994},
  organization={IEEE}
}

@inproceedings{gal2016dropout,
  title={Dropout as a bayesian approximation: Representing model uncertainty in deep learning},
  author={Gal, Yarin and Ghahramani, Zoubin},
  booktitle={international conference on machine learning},
  pages={1050--1059},
  year={2016},
  organization={PMLR}
}

@book{tarantola2005inverse,
  title={Inverse problem theory and methods for model parameter estimation},
  author={Tarantola, Albert},
  year={2005},
  publisher={SIAM}
}

@article{hastings1970monte,
  title={Monte Carlo sampling methods using Markov chains and their applications},
  author={Hastings, W Keith},
  year={1970},
  publisher={Oxford University Press}
}

@book{goodfellow2016deep,
  title={Deep learning},
  author={Bengio, Yoshua and Goodfellow, Ian and Courville, Aaron and others},
  volume={1},
  year={2017},
  publisher={MIT press Cambridge, MA, USA}
}

@article{foreman2012emcee,
  title={emcee: the MCMC hammer},
  author={Foreman-Mackey, Daniel and Hogg, David W and Lang, Dustin and Goodman, Jonathan},
  journal={arXiv preprint arXiv:1202.3665},
  year={2012}
}

@article{stuart2010inverse,
  title={Inverse problems: a Bayesian perspective},
  author={Stuart, Andrew M},
  journal={Acta numerica},
  volume={19},
  pages={451--559},
  year={2010},
  publisher={Cambridge University Press}
}

@article{fischer2003bayesian,
  title={Bayesian modelling of fusion diagnostics},
  author={Fischer, R and Dinklage, A and Pasch, E},
  journal={Plasma Physics and Controlled Fusion},
  volume={45},
  number={7},
  pages={1095--1111},
  year={2003}
}

@article{neal2011mcmc,
  title={MCMC using Hamiltonian dynamics},
  author={Neal, Radford M},
  journal={arXiv preprint arXiv:1206.1901},
  year={2012}
}

@article{marzouk2007dimensionality,
  title={Dimensionality reduction and polynomial chaos acceleration of Bayesian inference in inverse problems},
  author={Marzouk, Youssef M and Najm, Habib N},
  journal={Journal of Computational Physics},
  volume={228},
  number={6},
  pages={1862--1902},
  year={2009},
  publisher={Elsevier}
}

@article{karniadakis2021physics,
  title={Physics-informed machine learning},
  author={Karniadakis, George Em and Kevrekidis, Ioannis G and Lu, Lu and Perdikaris, Paris and Wang, Sifan and Yang, Liu},
  journal={Nature Reviews Physics},
  volume={3},
  number={6},
  pages={422--440},
  year={2021},
  publisher={Nature Publishing Group UK London}
}

@article{zhu2018bayesian,
  title={Bayesian deep convolutional encoder--decoder networks for surrogate modeling and uncertainty quantification},
  author={Zhu, Yinhao and Zabaras, Nicholas},
  journal={Journal of Computational Physics},
  volume={366},
  pages={415--447},
  year={2018},
  publisher={Elsevier}
}

@article{wang2021understanding,
  title={Understanding and mitigating gradient flow pathologies in physics-informed neural networks},
  author={Wang, Sifan and Teng, Yujun and Perdikaris, Paris},
  journal={SIAM Journal on Scientific Computing},
  volume={43},
  number={5},
  pages={A3055--A3081},
  year={2021},
  publisher={SIAM}
}

@article{abdar2021review,
  title={A review of uncertainty quantification in deep learning: Techniques, applications and challenges},
  author={Abdar, Moloud and Pourpanah, Farhad and Hussain, Sadiq and Rezazadegan, Dana and Liu, Li and Ghavamzadeh, Mohammad and Fieguth, Paul and Cao, Xiaochun and Khosravi, Abbas and Acharya, U Rajendra and others},
  journal={Information fusion},
  volume={76},
  pages={243--297},
  year={2021},
  publisher={Elsevier}
}

@article{hullermeier2021aleatoric,
  title={Aleatoric and epistemic uncertainty in machine learning: An introduction to concepts and methods},
  author={H{\"u}llermeier, Eyke and Waegeman, Willem},
  journal={Machine learning},
  volume={110},
  number={3},
  pages={457--506},
  year={2021},
  publisher={Springer}
}

@article{lakshminarayanan2017simple,
  title={Simple and scalable predictive uncertainty estimation using deep ensembles},
  author={Lakshminarayanan, Balaji and Pritzel, Alexander and Blundell, Charles},
  journal={Advances in neural information processing systems},
  volume={30},
  year={2017}
}

@article{seitzer2022pitfalls,
  title={On the pitfalls of heteroscedastic uncertainty estimation with probabilistic neural networks},
  author={Seitzer, Maximilian and Tavakoli, Arash and Antic, Dimitrije and Martius, Georg},
  journal={arXiv preprint arXiv:2203.09168},
  year={2022}
}

\end{document}